\documentclass[12pt,preprint]{aastex}

\begin{document}
\title{Motion properties of satellites around external spiral galaxies }

 \author{M. Azzaro, Isaac Newton Group of Telescopes, Apartado 321, 38700 S.C. de La Palma, Spain}
\author{F. Prada, Isaac Newton Group of Telescopes and Instituto de Astrof\'{\i}sica de Canarias,
v. L\'actea s/n, E-38200 La Laguna, Spain (Currently ``Ramon y Cajal'' fellow at IAA,
Granada, Spain)}
\author{C. M. Guti\'errez, Instituto de Astrof\'{\i}sica de Canarias,
v. L\'actea s/n, E-38200 La Laguna, Spain}

\begin{abstract}
We are analyzing a sample of closeby galaxy systems, each
comprising a bright isolated spiral and its satellites.
We find an excess (56\%) of prograde satellites over retrograde,
which basically holds for all angular displacements from the primary
major axis. Monte Carlo simulations show that interlopers and mixing
systems at different distances in the sample should not affect porcentages
sensibly.
\end{abstract}

\section{Introduction}
The currently accepted theories of galaxy formation rest on the
basic idea that large objects are formed by aggregation of low
mass elements. In this scenario, systems such as a massive primary
with its satellite galaxies are ideal gravitationally bound
environments where to investigate the mechanisms which
control mass aggregation in the Universe.\\
Due to the limited number of satellites per system
(Zaritsky et al. 1997, McKay et al. 2002, Prada et al. 2003),
a statistical approach is used here, following the method of
Zaritsky (Zaritsky et al. 1993): providing that the primaries are
selected in a consistent way, all the satellites are considered
as part of the same system, thus orbiting the same fictitious primary.\\
We selected a suitable sample of large isolated primary spirals
with their satellites, using data from the Sloan Digital Sky
Survey (SDSS) Early Data Release and Data Release 1. The selection
follows the criteria reported by Prada (Prada et al. 2003).\\
Here we show the preliminary results on the prograde/retrograde
ratio in our sample. We also compare our results with those
obtained by Zaritsky (Zaritsky et al. 1993, 1997), who uses a
sample of 57 primaries with measured rotation direction and 95
satellites. Through this contribution we call this sample ``ZS''.
ZS primaries velocity upper limit is 7000 km/s.

\section{The data}
The kinematics data on the selected sample
come both from the SDSS and from our observations.
We took receding velocities, absolute magnitudes
and positions for primaries and satellites from the
SDSS database; the morphological types of the primaries
have been taken from the Leda database and carefully
checked visually from the SDSS images, or devised visually
when not available in the literature.\\
Our sample includes 141 spirals with 200 satellites; the
primaries have been selected by absolute brightness $M_{B}$
to lay in the magnitude bin [-20.5 -19.5], and below 11000 km/s
in recessional velocity. This velocity limit was chosen
because, as by definition satellites are at least two
magnitudes fainter than their primaries, the satellites
of our faintest primaries will never be brighter than
$M_{B}=-17.5$. Given the apparent B magnitude limit of
SDSS (about 18.5, see Stoughton et al. 2002), $M_{B}=-17.5$ is
reached at 11000~km/s (h = 0.7).\\
Through an ongoing observational programme, we are measuring
the rotation direction of the primaries of the sample.
We measured the Sky Position Angle of the primaries from the SDSS
images and aligned the slit along the major axes.\\
The side-to-side Doppler shift was measured using the H$\alpha$
emission line around 6650 $\AA$, and comparing it with sky lines.
Our doppler shift detection has always been better than 10 $\sigma$.
Some of the galaxies show no detectable H$\alpha$ emission, while
for one object we could recover the rotation curve from the
literature (ngc2841, Afanasiev \& Silchenko 1999), yelding
a sample of 41 primaries with a satellite population
of 64 objects in total. If we want to make comparisons with
ZS, then we must match its velocity limit of 7000 km/s,
and then we are left with 23 primaries and 36 satellites.\\
Observations were carried out using the William
Herschel (4.2~m) and Isaac Newton (2.5~m) telescopes
at ING (La Palma), with the intermediate dispersion
spec\-tro\-graphs ISIS\- and IDS.
We performed a statistical analysis on the observed part
of the sample, which is shown in Figure~1 $a$ and $b$, and a
preliminary comparison with ZS.\\
For the statistical analysis, three ranges of angular separation
from the primary major axis are considered: $\pm 30^{\circ}$,
$\pm 45^{\circ}$, $\pm 45^{\circ}$, plus the total of
each subsample ($\pm 90^{\circ}$). The choice of selecting
satellites by their angular displacement from the primary
major axis was made to correlate the satellite motions
directly with the angular momentum of the primary.\\

\section{Conclusions}
Using the whole sample, we find around 56\% of prograde
satellites at all angular distances from the primary 
major axis. In the case of $v < 7000$ km/s, we find a
slightly higher porcentage (58\%) (Figure~1$a$), while the
ZS yelds 51\% over the total set and 50\% at small
angular distances from the primary major axis.\\
We find that the mean velocities of retrograde satellites are
nearly always much 

\begin{figure}
\plotone{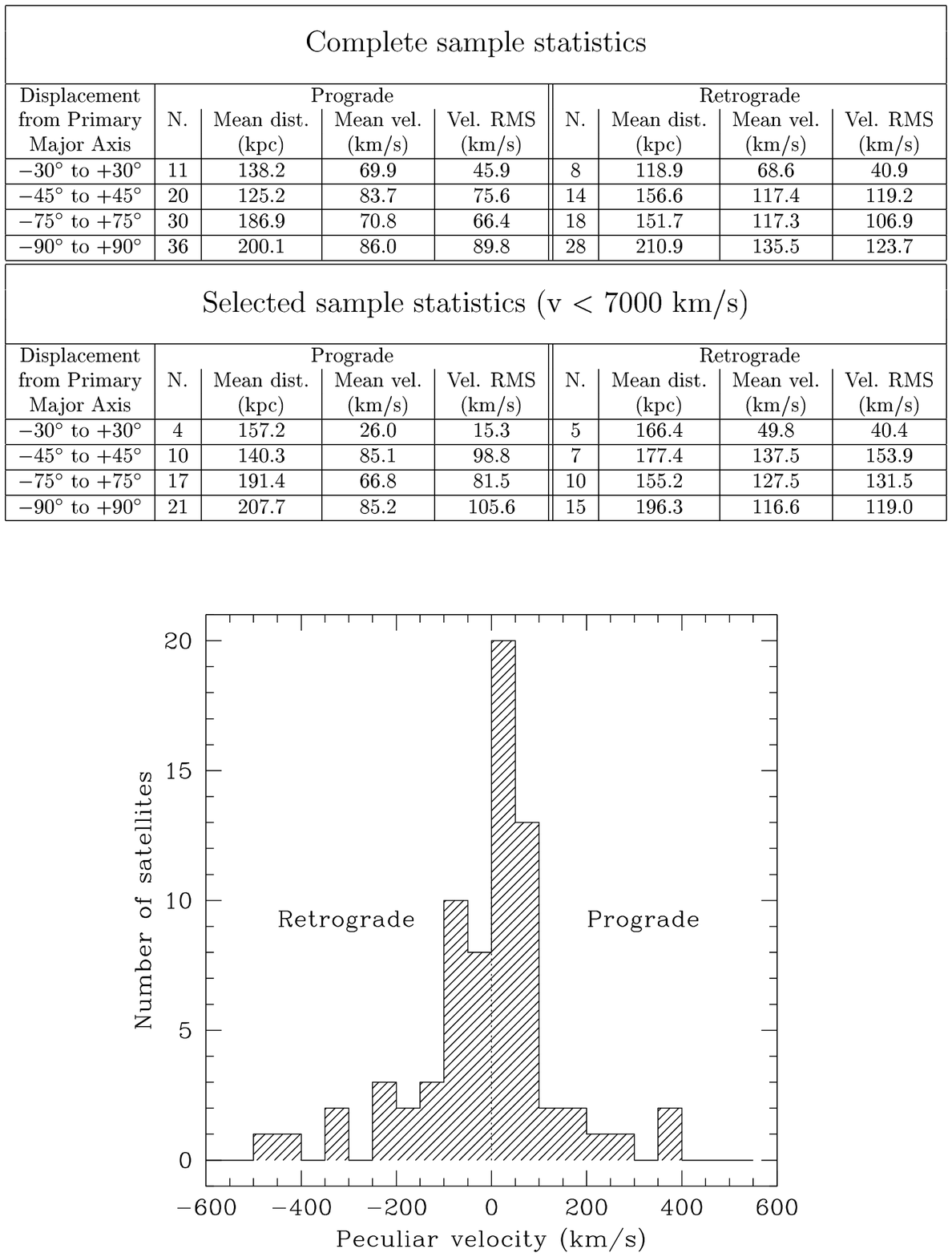}
\caption{Statistics of the prograde and retrograde populations.
The first panel shows our complete sample, the second panel
only the primaries from our sample with recessional velocity
below 7000 km/s like the Zaritsky sample.
The bottom histogram shows the distribution of satellites against
peculiar velocity.}
\end{figure}

\hspace*{-0.9cm} larger than those of prograde, while the mean
distances are usually comparable.\\
We also show a histogram of the distribution of prograde/retrograde
velocities in our sample (Figure~1$b$), where it can be seen that
the peak count is on the prograde side, but very close to zero
velocity.\\
Through Monte Carlo simulations, we estimated that the
contamination of interlopers (objects which are
field galaxies, not dinamically bound to the primary, but are
counted as satellites because of projection effects)
is small. Interlopers must be equally distributed between prograde
and retrograde, so their presence basically would ``push'' the
prograde/retrograde ratio towards the value of 50\%.
We estimate this effect to be of the order of 2\% for porcentages
around 60\% of prograde satellites.\\
Because we stack together systems of very different distances,
we could have a problem as magnitude completeness cannot be
fulfilled for all of them: some intrinsecally faint satellites
would be observed in closeby systems, but not in distant systems.
Through another Monte Carlo simulation, we are checking if this
problem could have any effects on the prograde/retrograde ratio 
and the results are promising, in the sense that the ratio seems
to be fairly insensitive to magnitude completeness and distance
mixing. In a future paper we will address these questions in more
detail.

\end{document}